\documentclass[sigconf,nonacm]{acmart}
\authorsaddresses{}
\usepackage{booktabs}
\usepackage{tabularx}
\graphicspath{{./}}

\begin{document}

\title[The Fly That Stopped]{The Fly That Stopped: Mushroom-Body-Inspired Habituation as a
Reward-Free Scheduling Prior for Autonomous Penetration Testing}

\author{Theodoros Moutesidis}
\affiliation{\institution{Independent Researcher}\city{}\country{}}
\email{}

\begin{abstract}
Autonomous security-testing agents can spend much of a fixed action budget
repeating earlier tool selections. We evaluate a reward-free scheduler inspired by
mushroom-body novelty processing in \emph{Drosophila}. It combines sparse state
encoding with decaying habituation counters over structural URL classes and tool
families. The counters penalize repeated clean or error outcomes without updating
weights from scalar reward. Four matched campaigns motivated this design by exposing
reward-accounting errors and tool-failure loops; reward-driven components did not
improve the tested primary outcomes over the reward-free MB condition. A
pre-registered pilot and two confirmatory stages then evaluated repeated
(tool, URL) selections. In the second confirmatory stage, 8 of 10 screened lab
targets remained measurable after two error-heavy slow-XSS exclusions. The
habituation-enabled scheduler lowered duplicate-action ratios in all 6 non-tied
target pairs (exact one-sided $p=0.015625$), with two ties; the largest reduction
was 51 to 18 duplicate steps within a 60-step budget. This is evidence for the
complete scheduler on the measurable budget-hold population, not an isolated
habituation ablation or a vulnerability-discovery gain. A complementary study on a
13{,}498-neuron MaleCNS-derived circuit (501{,}267 synaptic edges at weight
$\geq 5$) found no action selectivity from the five tested local-plasticity
approaches under a fixed readout; readout plasticity produced qualified positive
results in synthetic tasks without establishing a biological-topology advantage.
We report the population bounds, remaining input-integrity dependencies, and an
internal AI-assisted review protocol alongside the results.
\end{abstract}


\maketitle
\hypersetup{pdfauthor={Theodoros Moutesidis}}

\section{Introduction}\label{sec:intro}
An autonomous security-testing agent with a fixed action budget must decide when
to stop spending steps on familiar actions. In our scanner's lab harness, a static
baseline repeated earlier (tool, URL) selections on up to 51 of 60 steps. The scanner
already records tried actions and prioritizes coverage, but its frontier can
re-offer previously selected work. The baseline ranks those candidates directly;
MB additionally prefers untried pairs within a selected family. Neither policy
unconditionally forbids repetition (\S\ref{sec:mechanism}).

Reward-driven scheduling is one possible response. Four matched campaigns in this
program exposed difficulties with that response: recording defects could pay novelty
for repetition, error-prone tools distorted reward rankings, and protective
backpressure could resemble target failure. Correcting these problems removed much
of an apparent adaptive advantage. The remaining consistent difference was
repetition discipline, which the reward-free mushroom-body (MB) condition retained.

We therefore evaluate a scheduler inspired by mushroom-body habituation: a reduction
in response to familiar stimuli that recovers with time. Our implementation uses
sparse state encoding and a fixed readout for family scores, together with separate
decaying counters over structural URL classes, hypothesis families, and outcome
signatures. It does not train on scalar reward. That removes direct reward-value
updates but does not make it independent of observations: outcome classification,
finding-derived state features, and the frontier remain inputs whose integrity
matters.

The confirmatory chain tests the complete reward-free MB scheduler against the
static priority baseline. The pilot identified a repeated-selection effect and a
seed-dependence problem; confirmatory-1 located the effect on executions that exhaust
their budget; confirmatory-2 tested fresh targets selected for that execution
regime. Of 10 planned targets, 8 remained measurable after two slow-XSS exclusions.
Six pairs improved and two tied, with exact one-sided $p=0.015625$. We preserve
these population qualifications throughout. A separate connectome track examines
local plasticity in synthetic action-selection tasks; it does not train a fly
connectome to perform penetration tests.

\textbf{Contributions.} (1) An implementation of mushroom-body-inspired,
reward-free habituation in a cross-tool security-testing scheduler
(\S\ref{sec:mechanism}). (2) A pre-registered comparison showing lower measured
repetition for that scheduler on a selected, measurable lab population
(\S\ref{sec:chain}). (3) An account of reward-channel failure modes and the narrower
protection obtained by omitting reward-driven updates (\S\ref{sec:reward}).
(4) Bounded negative and qualified positive results for local versus readout
plasticity on a MaleCNS-derived circuit (\S\ref{sec:connectome}). (5) A documented
internal review workflow using pre-registration, frozen records, and separate
AI-assisted builder and reviewer sessions (\S\ref{sec:method}).

\section{Related Work}\label{sec:related}
\textbf{Computational-neuroscience ancestry.} Dasgupta, Stevens and
Navlakha formalized fly olfactory expansion coding as locality-sensitive
hashing~\cite{dasgupta2017flyhash}. Dasgupta et al.'s fly Bloom filter extends this
line to distance- and time-sensitive novelty detection~\cite{dasgupta2018flybloom},
with biological motivation from mushroom-body novelty and familiarity
responses~\cite{hattori2017novelty}. Our implementation adapts these ideas rather
than reproducing the full biological circuit or the original Bloom-filter update:
sparse code affects family scores, while decaying suppression operates over
engineered structural buckets. The contribution is their use and evaluation in a
security-testing scheduling loop. The unrelated Fruit Fly Optimization Algorithm
metaheuristic is outside this mechanism family.

\textbf{Scheduling and repetition control in security.} AFLFast reallocates
fuzzing energy using path frequency~\cite{bohme2016aflfast}; EcoFuzz and
T-Scheduler use bandit formulations~\cite{yue2020ecofuzz,luo2024tscheduler}, and
K-Scheduler uses graph centrality~\cite{she2022kscheduler}. AFuzz goes beyond
exact-key bookkeeping: it uses semantic scenario deduplication and embedding-based
diversity scheduling to reduce redundant investigations~\cite{agenticfuzzing2026}.
These are relevant precedents for spending a testing budget on less repetitive work.
Our emphasis is a fixed, reward-free habituation term with recovery and outcome
handling at the cross-tool family-selection layer; we do not claim to introduce
semantic deduplication or diversity scheduling.

Autonomous pentesting has been studied through reinforcement learning
(RL)~\cite{schwartz2019autopentest,rleval2024}, recommender
hybrids~\cite{sensors2025recommender}, curiosity-driven
exploration~\cite{curiosity2022pentest}, online-RL language-model
agents~\cite{pentestr1_2025}, and LLM orchestration~\cite{deng2024pentestgpt}.
Repetition control in this literature is not uniformly a hard mask. Yang and Liu,
for example, combine action-coverage history with state-dependent learned fusion
weights and an auxiliary coverage loss~\cite{curiosity2022pentest}. Our distinction
from that approach is the absence of reward-trained scheduler weights and the use
of fixed, decaying outcome counters. Our experiments compare the implementations in
\S\ref{sec:matched}, not the full range of these published systems.

\textbf{Habituation and biological analogies.} Marsland, Nehmzow and Shapiro
applied habituation to novelty detection in mobile robotics
\cite{marsland2000habituation,marsland2005online}; these are robotics antecedents,
not evaluations of security defenders. Artificial immune systems provide a separate
biological analogy for security detection~\cite{forrest1994selfnonself,hofmeyr2000ais}.
Stirewalt and Gebremedhin's adversarial-habituation poster studies exploiting a
defender's habituation~\cite{stirewalt2025habituation}, whereas our counters regulate
the testing agent's own choices. Outside security, the public
\texttt{connectome-fighter} repository explores agents for FightingICE using
MaleCNS~\cite{connectomefighter}. Neither a biological analogy nor a shared dataset
establishes that biological topology is responsible for an observed benefit.

\textbf{Experimental methodology.} Design and reporting guidance exists in
security research~\cite{coopamootoo2018toolkit}. Recent work includes a
pre-registered insider-threat simulation~\cite{hbee2026}, an executable
pre-registration contract for an agent-assurance evaluation~\cite{execontract2026},
and a template for experiments with AI agents~\cite{preregtemplate2026}. Our
program also includes a companion verifier-stage
ablation~\cite{moutesidis2026verifier}. We contribute a concrete internal workflow
for frozen analysis, separate-session adversarial checks, and explicit corrections.
We make no field-wide priority claim for pre-registration or independent review.

\section{The Corrupted-Reward Problem}\label{sec:reward}

Reward-driven scheduling presumes a reward channel that measures the target. In an
autonomous security agent, we found that channel to be attack surface --- corrupted
in practice by the agent's own tooling, its protective machinery, and its evidence
pipeline. We catalogue five corruption classes, observed within the research program, with the accounting exhibits that revealed them.

\textbf{(1) Recording artifacts pay real reward.} Tools that take non-URL arguments
recorded an empty URL, which defeated repeat-detection and paid novelty on every
re-execution. The natural experiment that exposed it: two conditions exhibiting the
\emph{same} pathological behavior --- repeatedly re-running one tool into a
100\%-error pit --- scored $+44.4$ (the pit tool recorded an empty URL) versus $-8.1$
(the pit tool recorded its URL). Identical behavior, opposite signs, and the learning
conditions of that campaign had already trained on the corrupted signal; behavior
contamination has no offline fix, and the entire campaign's learning comparison had
to be re-run.

\textbf{(2) Error pits pay arbitrage.} With recording repaired, the contextual
bandit's apparent clean-reward lead over both MB conditions decomposed into pit
\emph{substitution}: it escaped a skip-penalty pit by relocating into an error-prone
tool's pit (14--21 error steps per cell versus MB's 4--9), and a skip-neutral
counterfactual re-scoring flipped every adaptive-vs-adaptive pairing. A reward
ranking that inverts under a one-line accounting change is not a ranking; the
adjudication campaign (\S\ref{sec:matched}) settled it only after the error class was
eliminated outright.

\textbf{(3) The agent's own protection reads as punishment.} The scanner's rate
limiter escalates under pressure and answers with synthetic 503 responses that are,
at the reward interface, indistinguishable from target failure --- so a
reward-learner is trained to avoid surfaces \emph{because the agent protected them}.
The intended interface distinguishes an explicit \texttt{throttled} outcome
from a target error and gives it neutral treatment. The habituation implementation
ignores outcomes carrying that flag. This protection is conditional on correct
propagation: the frozen live harness used for the confirmatory chain did not pass
an explicit throttle flag to the controller. Its analysis exports contain no
reported throttle events, but that disclosure is not a proof of end-to-end immunity.

\textbf{(4) The evidence pipeline can fabricate.} LLM-based finding enrichment was
observed writing analyst verdicts into stored findings that contradicted the recorded
evidence. Any reward derived from enriched fields inherits this; our reward-feedback
contract is enrichment-blind by property test. The value of the deterministic
governance layer that stands between raw claims and reward is measurable: on a
local 104-benchmark pool, governance downgraded 6.0\% of raw findings and
compressed 21.3\% of critical-band claims, with zero upgrades; independent
live-replay verification of 101 of 350 findings confirmed 101 true positives and 0
false positives (the replayed subset is higher-signal-skewed; we make no accuracy
claim about the unverified rows).

\textbf{(5) The harness leaks into the metric.} One campaign's error-step metric
partly measured the execution environment, not the target: a missing harness
dependency produced ${\sim}777$ error rows for one tool that never reached any
target, uniformly across conditions. Validity gates on later campaigns (zero error
rows campaign-wide, empty-URL audits) exist because of this class.

\begin{figure*}
  \centering
  \includegraphics[width=.82\textwidth]{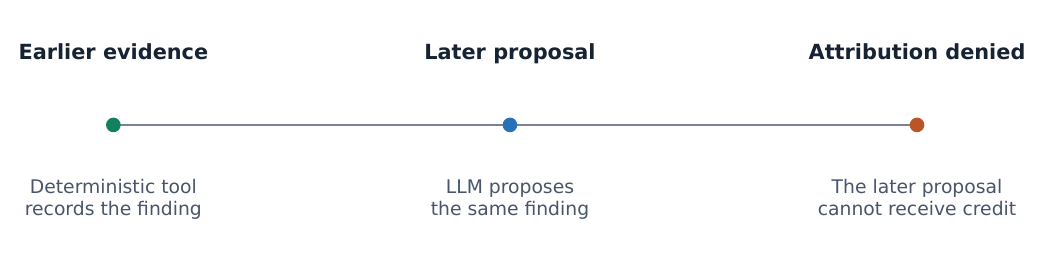}
  \Description{Timeline showing a deterministic finding at t0, an LLM re-proposal at t1, and an attribution-denied stamp from the temporal join.}
  \caption{Provenance as a temporal join: the deterministic floor finds a
  vulnerability at $t_0$; the LLM proposes the same candidate at $t_1$; attribution
  is denied. This rule falsified our own headline metric once (\S\ref{sec:method}).}
  \label{fig:provenance}
\end{figure*}

These observations motivate removing a source of dependence rather than claiming
complete robustness. A scheduler with no scalar-reward update cannot be trained by
a fabricated scalar reward. It can still respond incorrectly to erroneous outcome
labels or state features, and tool failures can waste its budget directly. We did
not evaluate adversarial manipulation of the remaining inputs. Figure~\ref{fig:provenance}
illustrates the separate requirement to attribute results to the process that
actually produced them.

\section{Mechanism: Reward-Free Habituation in an MB-Inspired Scheduler}\label{sec:mechanism}

\subsection{Setting and selection unit}
The scanner builds a candidate frontier from deterministic coverage planning.
A candidate contains a tool, URL, arguments, static priority, and hypothesis family.
The MB controller first selects a \emph{family}, then a candidate within it; it does
not assign each candidate its own neuron. Production integration records shadow
choices; the lab harness instead lets each controller drive selection directly.
The experiments in \S\ref{sec:chain} use deterministic scanner lanes, with no LLM
selecting actions. Conditions share the state builder, frontier, budgets, and execution plumbing.
MB and LinUCB use the same within-family rule; the priority baseline instead ranks
all candidates directly by coverage obligation, static priority, and candidate ID.
It does not apply MB's explicit untried-first preference. Thus the comparison is
between complete selection policies, not a one-factor habituation ablation.

\subsection{State encoding and fixed readout}
Fifteen scan-state features are clamped and normalized to $[0,1]$, then concatenated
with fifteen measured-value indicators. Missing values have value and indicator
zero; a measured zero has indicator one. The features describe progress, execution
outcomes, findings, and coverage (Appendix~\ref{app:features}); they are
\emph{scan-level}, not separate candidate vectors.

A seeded, fixed projection maps the 30 inputs to 8{,}192 Kenyon-cell-like units.
Each unit samples six distinct inputs with weights $\pm1$. The encoder retains up
to 164 units with positive activation (approximately 2\%); if none are positive,
the code is empty. A fixed Gaussian readout, initialized with standard deviation
0.05 and seed one greater than the projection seed, supplies a family score $Q_f$
by averaging its weights over active units. These weights are not trained in
\texttt{mb\_static}. The design is inspired by sparse expansion coding; we do not
attribute the confirmatory result specifically to that encoder.

\begin{figure*}
  \centering
  \includegraphics[width=.95\textwidth]{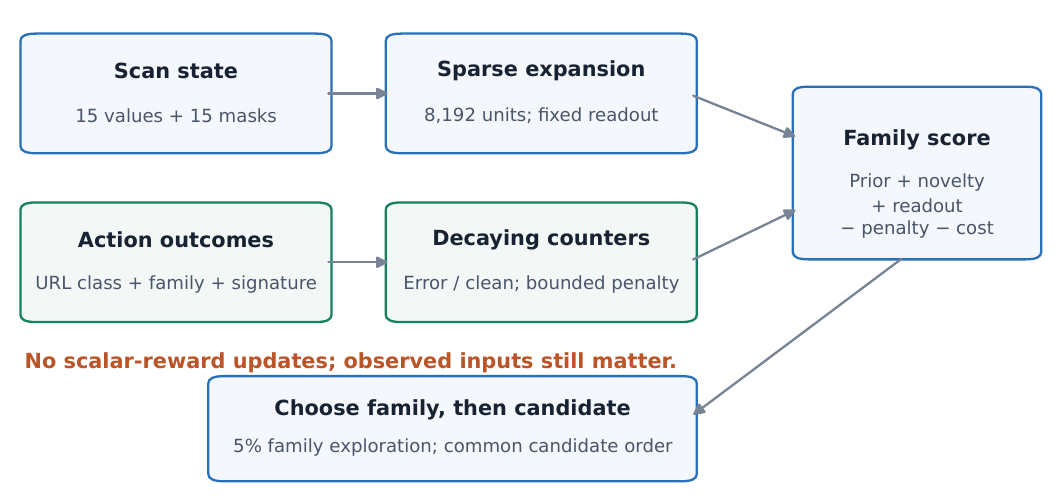}
  \Description{Separate scan-state encoding and structural outcome-counter paths combine in a family score. Family selection precedes the common within-family candidate rule.}
  \caption{Two scoring paths in the reward-free MB scheduler. Sparse state encoding
  supplies a fixed readout term. Structural outcome counters supply a separate
  habituation penalty. Neither path trains on scalar reward; both depend on the
  integrity of their observed inputs.}
  \label{fig:mechanism}
\end{figure*}

\subsection{Outcome counters and recovery}
Counters are keyed by (surface class, hypothesis family, outcome signature).
The URL class comprises kind (API, asset, or page), nonempty path-segment count
capped at four, and query-string presence. Kind uses fixed URL heuristics:
\texttt{/api}, \texttt{graphql}, or JSON/XML extensions denote API surfaces;
a fixed static-file extension list denotes assets; other URLs are pages.
Thus different URLs can share a counter, although the primary repetition metric
uses exact (tool, URL) identity (\S\ref{sec:metric}).

Before each selection, existing counters decay as $H_k\leftarrow0.9H_k$;
values below 0.01 are removed. After an observed action, the matching error or clean
counter is incremented by one. Explicitly flagged throttles and skipped actions do
not increment counters. For a non-error, non-skipped action, a positive reported
finding count yields a hit, which also does not increment a counter; otherwise the
outcome is clean. A hit does not erase accumulated suppression.

For surface $s$ and family $f$, the bounded penalty is
\begin{equation}
 B_{s,f}=1-\exp[-0.7(H_{s,f,\mathrm{error}}+0.5H_{s,f,\mathrm{clean}})].
\end{equation}
This is a \emph{score penalty}, not an independently sampled suppression
probability. The family penalty $B_f$ is its mean over distinct surface classes
represented among that family's current candidates. State resets between scan
conditions; there is no cross-target habituation transfer.

\subsection{Family score and action choice}
Let $P_f$ be the maximum candidate static score divided by ten and clipped to
$[0,1]$, plus 0.5 if any candidate has a coverage obligation. Let $N_f$ be the
fraction of candidates whose exact (tool, URL) key is absent from the scan's
execution-history tried set. Let $C_f$ be the mean nonnegative estimated cost,
divided by ten and capped at one. The activation is
\begin{equation}
 A_f=P_f+0.5N_f+Q_f-B_f-0.1C_f.
\end{equation}
Soft lateral inhibition subtracts $0.25\sum_{g\ne f}\max(0,A_g)$ before ranking.
With probability 0.05, selection explores uniformly among nonempty families;
otherwise it takes the highest-ranked family, breaking score ties by family name.
The exploration draw is seeded by the configured seed and current state hash.
That hash includes scan ID, step index, and measured features; fixing the nominal
seed does not force identical exploration draws across different scan IDs.

Within the selected family, the common lexicographic rule prefers coverage
obligations, then untried pairs, then higher static priority, then candidate ID.
Exploration therefore gives each available \emph{family} a route to selection; it
does not guarantee a nonzero probability for every candidate. Novelty $N_f$ uses
exact tried-pair history, whereas habituation generalizes over structural buckets.
These are distinct mechanisms. In the live harness, all candidates come from the
coverage frontier and retain the default estimated cost of one, making the coverage
boost and cost term constant across available families.

\subsection{What reward-free means}
The tested \texttt{mb\_static} configuration has learning disabled: no scalar reward
updates its projection, readout, or counters. It is not observation-free. Finding
counts distinguish hits from clean outcomes, and the fixed-readout state vector
includes finding totals and validation/severity ratios. Consequently the scheduler
can still be affected by altered findings, outcome classification, or frontier
construction. Reward-free here means absence of reward-driven parameter updates,
not general immunity to misleading feedback.

\subsection{Relationship to existing repetition controls}\label{sec:residue}
Frontier construction merges simultaneously proposed candidates with the same
identity. Tried-action history then affects coverage planning and within-family
preferences, but these mechanisms do not prohibit all repeat selections over time.
The measured comparison asks whether the full MB scheduler reduces the remaining
repeat count relative to the existing priority discipline. It does not establish
that a complete visited-set could not prevent those repeats, nor that all measured
repeats are semantically useless. A counter-only or habituation-disabled MB ablation
would be required to isolate individual components.

\section{Matched Comparisons of Scheduler Configurations}\label{sec:matched}

\subsection{Design}

Five controller conditions share the state builder, candidate frontier, budgets,
and feedback plumbing, with selection-policy differences described in
\S\ref{sec:mechanism}: two static baselines (the production priority discipline \texttt{pq} and the
legacy scheduler, which proved decision-identical to it absent an in-scan fatigue
signal), a LinUCB contextual bandit, reward-free \texttt{mb\_static}, and
\texttt{mb\_plastic} (habituation plus eligibility-trace plasticity). Four live
campaigns ran on local benchmark labs under progressively stricter validity gates:
run 1 (two targets, single seed, exploratory), run 2 (five targets $\times$ three
seeds $\times$ five conditions, 75 cells, fresh labs, shuffled condition order), run
3 (same shape, first valid reward accounting), and run 4 (adjudication: both known
tool pits eliminated, zero error rows campaign-wide, per-URL availability semantics
fixed). The subsequent pre-registered comparisons used evidence yield and objective
flag capture as primaries; run 1 was exploratory.

\subsection{The arc: an apparent adaptive advantage, and its decomposition}

Run 1 suggested an adaptive advantage: adaptive conditions delivered
$6\times$ the graded evidence of static baselines on an error-heavy target, and MB
recorded a duplicate ratio of 0.00 on one target. Review narrowed the interpretation:
the evidence concerned generic hygiene, no condition reached the benchmark's
vulnerability class, and all adaptive conditions repeatedly selected a tool that
failed on every execution. The bandit spent over half its budget on that tool and
MB conditions spent 40--47\%. The yield difference therefore reflected allocation
around a tool failure rather than demonstrated investigation quality.

Run 2 repeated the yield gap across more pairs (11--12 of 12 error-heavy pairs),
but adaptive-to-adaptive evidence yields tied in 13 of 15 pairings and flag capture
was zero in all 75 cells. The recording defect in \S\ref{sec:reward} invalidated
its reward comparison. Run 3 repaired that accounting and refuted the
pre-registered learning hypothesis under the recorded reward; however, its ranking
changed under the skip-neutral counterfactual, motivating a further adjudication.

Run 4 removed both known tool-failure loops and recorded zero error rows.
\textbf{MB-static, MB-plastic, and LinUCB tied on evidence yield in all 15 matched
cells for each of the three pairwise comparisons.} Static baselines recovered much
of the earlier gap: adaptive-to-static yields tied on two of five targets, with all
five conditions identical on one. Across the four campaigns, the apparent
$6\times$ adaptive advantage was largely explained by routing around tool failures.
Flag capture remained zero in each of the three 75-cell campaigns.

\subsection{Findings that remained after correction}
Repeated-selection control remained observable in reward-free \texttt{mb\_static}.
Reward-driven components did not improve the tested primary outcomes over that
condition. Section~\ref{sec:chain} evaluates the complete configuration against
\texttt{pq}; it does not isolate each scoring term.

\begin{figure*}
  \centering
  \includegraphics[width=.95\textwidth]{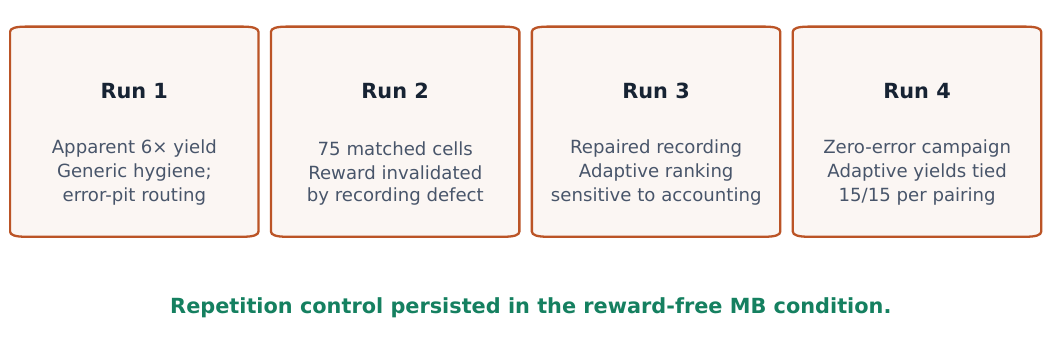}
  \Description{Four campaign tiles with their corrected verdicts, resolving into one surviving advantage: duplicate discipline.}
  \caption{Four matched campaigns and five conditions: accounting corrections and
  the persistent repetition-control observation. Each
  tile carries both the initial interpretation and its subsequent correction.}
  \label{fig:arc}
\end{figure*}

Two additional mechanism observations were conditional on their test regimes.
LinUCB acquired avoidance of skipped actions roughly $3\times$ faster than
eligibility-trace plasticity under an identical feedback signal, with parity under
weak penalties. In synthetic switch worlds, the bandit's apparent re-adaptation
advantage instead reflected approximately 50\% successful lock-in: per-seed
post-switch reward-producing action rates clustered at 1.0 or 0.03. MB adaptation
was more consistent but weak. These observations motivate further study of credit
assignment and recovery; they do not establish a general superiority ordering.
Figure~\ref{fig:arc} summarizes the four live campaigns.

\section{The Confirmatory Chain}\label{sec:chain}

\subsection{Primary metric and frozen implementation}\label{sec:metric}
For each scan condition, the harness maintains a fresh set $S$ of selected
(tool, URL) keys, using the first 12 hexadecimal characters of SHA-1 over the
literal string \texttt{tool|url}. No additional URL normalization, role, payload,
or argument fields enter this metric. After an action reaches the metric-update
point, it increments the duplicate count if its key is already in $S$, then adds
the key. Tool success, error, and skipped statuses are counted alike when they
reach that point. An exception before metric update does not add a duplicate;
its loop position can still contribute to the denominator.

The reported ratio is $D/N$, where $D$ is that repeat count and $N$ is
\texttt{steps\_used}: the last selected action's one-based loop index, capped at 60.
The fingerprint/crawl bootstrap is excluded. Ordinarily $N$ equals the number of
selected actions; under harness exceptions it can also include intervening failed
iterations. A frontier-drained cell can therefore have $N<60$. Different URLs of
the same shape do not count as repeats, while the same tool and URL with different
arguments do. We interpret this endpoint as selection repetition, not an exhaustive
measure of redundant security work. It is distinct from the controller's tried set,
which is read from persisted execution records.

The confirmatory-2 implementation was frozen at commit \texttt{b3f7bf6b1bc1};
its relevant controller, feature, and metric files were checked against that commit
when preparing this paper. The confirmatory analysis uses paired target-level differences
$D_{\mathrm{pq}}/N_{\mathrm{pq}}-D_{\mathrm{MB}}/N_{\mathrm{MB}}$, drops zero
differences, and applies a one-sided exact Wilcoxon signed-rank test at
$\alpha=0.05$, requiring at least five non-tied targets.

\subsection{Pilot (pre-registered)}

\begin{figure*}
  \centering
  \includegraphics[width=.9\textwidth]{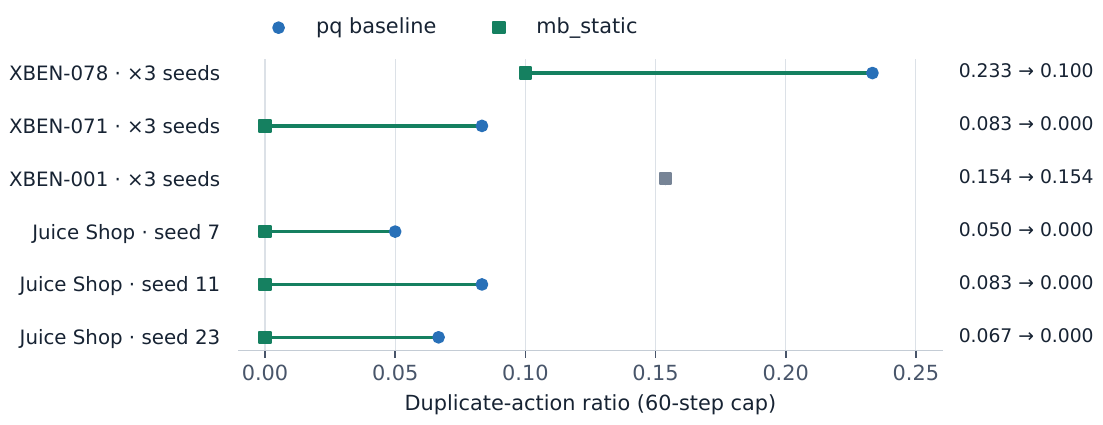}
  \Description{Horizontal paired plot of pilot duplicate-action ratios, with seed-identical XBEN pairs collapsed.}
  \caption{Pilot: paired duplicate-action ratios; 9 of 9 non-tied pairs fell, exact
  $p = .002$. Seed-identical XBEN lines are collapsed with multiplicity tags, so
  the effective-$n$ annotation is structural: XBEN seeds are not independent
  replicates (a $\times 3$ line is one line) --- conservative target-level
  $p = .125$, intermediate $p = .031$.}
  \label{fig:pilot}
\end{figure*}

Four lab targets (XBEN-001, XBEN-071, XBEN-078, OWASP Juice Shop) $\times$ three
seeds $\times$ \{\texttt{pq}, \texttt{mb\_static}\}, 60-step cells, live scanner
lanes, pre-registered primary: \texttt{mb\_static} produces lower duplicate-action
ratios. The seed-level test rejected the null: 9 of 9 non-tied (target, seed) pairs fell, exact one-sided
$p = 0.001953125$ (12 pairs, 3 ties dropped; Figure~\ref{fig:pilot}). The required
annotation travels with that number: XBEN seeds are not independent replicates for
this metric --- on the XBEN targets, duplicate ratios were identical across seeds ---
so conservative collapse readings are $p = 0.125$ (target-level, not significant) and
$p = 0.031$ (intermediate). Baseline repeat counts on the non-tied pairs were 3--14 per 60-step budget;
these are baseline counts, not avoided-step deltas. The
pre-registered consequence: any confirmatory run uses the target as the replicate
unit.

\subsection{Confirmatory-1: the effect has a surface}

Nine fresh targets $\times$ one seed $\times$ both conditions. Verdict (frozen rule):
\emph{insufficient non-tied targets} --- 7 of 9 pairs tied exactly, all
frontier-drains that terminated at their deterministic-probe depths, leaving $T = 2$
non-tied pairs against a pre-registered minimum of 5; the minimum-sample gate prevented a confirmatory decision. Both non-tied pairs exhausted the budget, favored MB, and met the
pre-registered absolute-repeat discriminator (XBEN-039: $0.233 \rightarrow 0.100$; XBEN-061:
$0.533 \rightarrow 0.000$ --- \texttt{pq} spent 32 of 60 steps on duplicates, the
largest baseline repeat count in the program at that point; confirmatory-2 later matched
and exceeded it). Zero discordant. The run's lesson is a boundary, not a null:
the observed between-condition difference occurred on \emph{budget-hold} executions --- targets whose frontier
outlives their budget --- and generalization of the pilot effect is bounded by the
prevalence of that surface. A descriptive observation for future work:
drain-path duplication was condition-invariant (6--8 absolute duplicate steps in
every tie).

\subsection{Confirmatory-2: confirmed on the budget-hold population}\label{sec:confirm2}

A pre-registered screen over a 104-benchmark local pool identified budget-hold
targets (10 holds across 19 attempts, stopped by its pinned cap); the frozen design
took a 10-target final set at one seed, \{\texttt{pq}, \texttt{mb\_static}\}, 60-step
cells. Two targets were dropped by a pinned 5-point validity checklist (both wedged
the harness --- the record's hypothesis is a step-timeout escape; both are
error-heavy slow-XSS targets), reducing the measurable population to 8 --- a
mechanically caused amendment that the campaign record documents as outcome-blind
and countersigned by the separate-session reviewer before result inspection.

\begin{figure*}
  \centering
  \includegraphics[width=.9\textwidth]{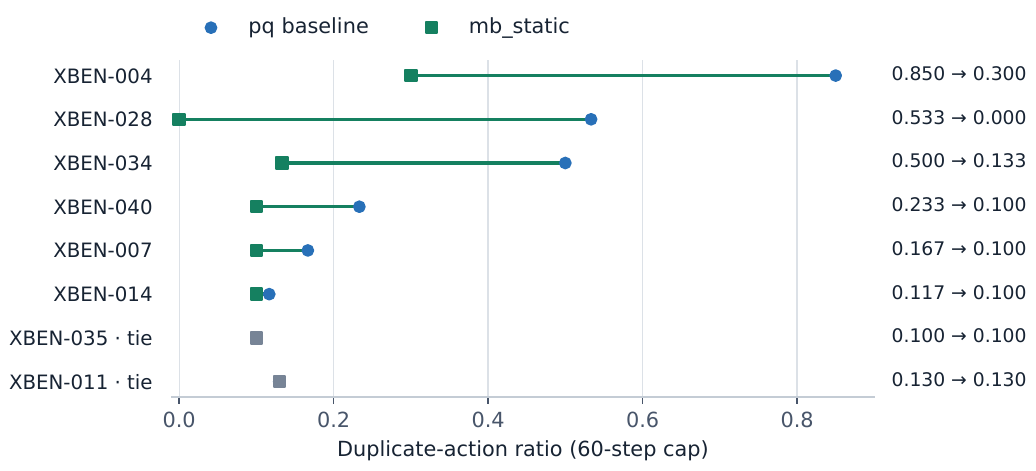}
  \Description{Horizontal paired plot of confirmatory-2 ratios, with six improvements and two grey ties.}
  \caption{Confirmatory-2: fresh budget-hold targets, one pair per target on the
  measurable population. Six of six non-tied pairs fell, exact $p = .0156$;
  population 8 of 10 measurable --- both drops error-heavy slow-XSS (the
  family-composition bound travels); suppression, not elimination: one hard-zero
  in six; both ties labelled explicitly.}
  \label{fig:confirm2}
\end{figure*}

\textbf{Result: 6 of 6 non-tied pairs concordant, $\texttt{mb\_static} <
\texttt{pq}$, exact one-sided Wilcoxon $p = 0.015625$, zero discordant}
(Figure~\ref{fig:confirm2}). The pre-registered absolute-repeat discriminator passed:
every non-tied pair had \texttt{pq} recording $\geq 7$ absolute duplicate steps (XBEN-004:
51, XBEN-028: 32, XBEN-034: 30, XBEN-040: 14, XBEN-007: 10, XBEN-014: 7), with both
conditions budget-exhausted in all six. Both ties are named: XBEN-035 is the
program's first true budget-hold tie ($0.100 = 0.100$), and XBEN-011's hold status
itself did not reproduce (both cells frontier-drained where the screen held).
Mechanism grading: \emph{suppression, not elimination} --- one hard-zero in six
(XBEN-028: $0.533 \rightarrow 0.000$); the remaining five pairs compress without
vanishing ($0.85 \rightarrow 0.30$, $0.50 \rightarrow 0.13$, $0.23 \rightarrow 0.10$,
$0.17 \rightarrow 0.10$, $0.12 \rightarrow 0.10$).\footnote{The frozen export contains $p=0.015625$ and the literal verdict
\texttt{incomplete\_pairs}; its SHA-256 begins \texttt{7f2d7300}. The campaign
record adjudicates the result on the amended eight-target population, while the
analyzer's verdict retained the original ten-target count gate. The record also
says the $p$ value was not computed, which conflicts with the preserved export.
We report the export as it exists and independently reproduce its statistic from
the eight pair summaries; a subsequent wording-reconciliation memo remains
unratified. No raw data, export, or historical verdict was rewritten for this paper.}

\subsection{Recorded population-qualified claim}

\begin{quote}
On fresh budget-hold lab targets that remained measurable (8 of 10; two error-heavy
slow-XSS targets dropped for harness-pathology wedges --- the population
under-represents that family), habituation-only mb\_static produced lower
duplicate-action ratios than pq in 6 of 6 non-tied pairs (exact one-sided p=0.0156,
target-level replication), with two named ties; the mechanism is suppression, not
elimination (one hard-zero in six).
\end{quote}

Here the record's ``habituation-only'' names the \texttt{learn=False} MB
configuration; it does not denote an ablation holding every other scoring term
constant. The pilot effective-$n$ and confirmatory-1 surface-prevalence
qualifications apply (\S\ref{sec:discussion}).

\subsection{Interpretation}
The chain progresses from a pilot association with a replication-unit caveat,
through a largely tied broader target set, to a confirmatory comparison within a
screened execution regime. The six improving confirmatory-2 pairs have baseline
repeat counts of 7--51 and avoid 1--33 repeats; only XBEN-028 reaches zero.
Appendix~\ref{app:pairs} lists all eight pairs. The test establishes a difference
between the complete scheduler configurations on this population. It does not
separate habituation from exact-pair novelty, fixed readout, cost scoring, or
family-level exploration, and it does not demonstrate increased vulnerability yield.
The small sample, selected targets, one seed per confirmatory target, and correlated
exclusions limit generalization.

\section{Local and Readout Plasticity on a Connectome}\label{sec:connectome}

A separate track asked whether biologically local plasticity on a real connectome
could support action selection in synthetic tasks. This is a complementary
mechanism study, not the implementation used in the scanner or a penetration-testing
capability evaluation. The results distinguish the tested fixed-readout and
readout-plastic configurations within a specified circuit and task envelope.

\subsection{Substrate and design}

From the MaleCNS connectome (v1.0)~\cite{malecns2026} we extracted the mushroom-body and
central-complex circuit: 13{,}498 neurons and 501{,}267 synaptic edges at weight
$\geq 5$, frozen as a hashed artifact with pinned seeds. Four arms share one training
computation and differ only in wiring: the real topology (A), a degree-preserving
shuffle (B), a random graph (C), and a non-graph engineered reference (D, the
comparability anchor --- byte-identical across campaigns by construction). Synthetic
worlds (a decoy-reward world and a mid-episode-switch world) supply delayed,
decoy-laden credit at 10 seeds $\times$ 30 episodes $\times$ 300 steps per cell; the
pre-registered primary is episodes-to-criterion at a frozen threshold (a 5-episode
rolling reward mean $\geq 0.5\times$ optimal), with topology separation
($A > B \wedge A > C$) as the machine-readable secondary.

\subsection{Five local-plasticity approaches under a fixed readout}\label{sec:ladder}

The five approaches had pre-registered falsification criteria, frozen control
configurations with byte-identity gates, and separate-session recomputation. In
these synthetic tasks, the ``paying'' action is the action that produces reward.

\textbf{Action-conditioned three-factor credit (A1).} No tested variant followed a
swap of the paying action label on either toy task. An ablation grid varied five
baseline conventions, exploration rates from 0 to 0.8, horizons from 400 to
8{,}000 steps, seven seed sets, and credit budgets up to $50\times$ the baseline.
The tested variants still failed the criterion, even where attainability analysis
showed that the paying configuration was reachable within the parameter bounds.

\textbf{State-conditioned credit (A2).} Adding varying state features that identify
the paying action did not make uniform credit label-selective in the mechanism
test. The connectome campaign produced 162 output files and preserved all 80 base
world files byte for byte. Varying the input roughly halved total plastic change
and shifted reward distributions, but no cell crossed the performance criterion.

\textbf{Advantage gating (A3).} Review found that one initial falsification cell
allowed only a single reachable action choice, so it could not test selectivity.
After correcting that test and checking reachable action sets, gated, ungated, and
oracle-attributed variants all failed to establish selectivity in the tested cells.

\textbf{Weight decay.} Decay reduced stale credit mass by factors of 6.9--13.3
without improving selection. The eventual selected action was determined early by
credit deposited during exploratory bursts, whereas decay affected only about 7\%
of that mass per burst. Removing older credit did not resolve this immediate
competition.

\textbf{Commit gating (K4).} Gating concentrated positive credit on the paying
action: median scheduling acceptance was 0.992, the accepted set contained about
one action, and the reported rival-credit blocking ratio was approximately
12.7:1. Despite this improvement in credit attribution, none of 60 graph cells
crossed the criterion, and late selection of the paying action remained at chance.
Improved attribution alone was therefore insufficient in this configuration.

Across these approaches, reducing a measured source of credit contamination did
not reliably produce action selectivity. This negative result is restricted to the
tested rules, parameter ranges, substrate, fixed readout, and synthetic worlds.
It is not a general impossibility result for local learning or connectome-based
control. One remaining hypothesis is a mismatch between the states represented
during reward-producing bursts and those encountered at later selections; this
explanation remains untested.

\subsection{Qualified positive results from readout plasticity}
Moving plasticity from graph edges to the readout, while freezing the graph,
produced qualified positive results. In the toy-task pair, the varying-input
configuration passed the prescribed test in both deciding cells. Replacing graph
propagation with a hop-free encoder at the same sparsity reduced the deciding test
outcomes from 4/4 to 1/4. That single-seed ablation is indicative only. A separate
re-implementation without importing the original implementation also reproduced
the pass. Alternative fixtures failed the full acceptance conjunction, limiting
the claim to the tested distribution rather than reliable success in every cell.

The campaign verdict remained \texttt{inconclusive\_setup}:
no cell reached the criterion, and the best sustained rolling mean reached 46\%
of the threshold. Relative to the edge-plastic predecessor with matched features,
graph-arm final-window median reward increased by 30--69 units, while the non-graph
reference changed by exactly zero. Late selection of the paying action increased
from below chance to approximately chance. Real topology did not separate from
shuffled or random wiring (secondary $p=0.28$--$0.95$; ordering changed between
worlds). These comparisons establish no biological-topology advantage.

\subsection{Sensitivity to exploration}
A pre-registered probe changed only exploration, from $\varepsilon=0.05$ to zero.
It produced nine criterion crossings compared with none under exploration. The
best sustained performance increased from 46\% to 168\% of the threshold, while
medians did not change. This intervention implicates interference from exploratory
actions in the tested regime, but the benefit was concentrated in successful early
lock-in; many executions remained near chance.

\subsection{Interpretation across the two tracks}

\begin{figure*}
  \centering
  \includegraphics[width=.9\textwidth]{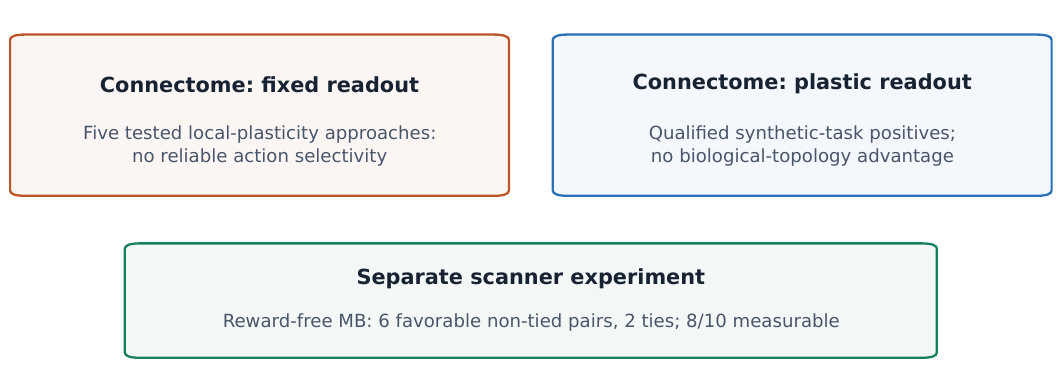}
  \Description{Summary of bounded connectome results and the separate live-scanner scheduling result.}
  \caption{The experiment map separates fixed-readout negative results, qualified
  readout-plasticity results, and the scheduler comparison. No biological-topology
  advantage was demonstrated.}
  \label{fig:map}
\end{figure*}

The tested local-plasticity approaches did not induce action selectivity under the
fixed readout. Readout plasticity produced qualified positive results, and removing
exploration enabled some criterion crossings, but biological topology did not
separate from shuffled or random wiring. These findings do not establish a unique
trainable locus for other rules or environments. They motivated evaluating the
simpler, reward-free scheduler separately. Figure~\ref{fig:map} summarizes the
outcomes while keeping the two experimental tracks distinct.

\section{Methodology: Pre-registration and Adversarial Review}\label{sec:method}

The program uses pre-registration and a separate-session review workflow to make
analysis choices and corrections inspectable. Related approaches appear in the
companion verifier study~\cite{moutesidis2026verifier} and other agent research
\cite{hbee2026,execontract2026,preregtemplate2026}. We present this workflow as a
practical contribution, without claiming that earlier security research lacks such
methods.

\textbf{Pre-registration with frozen analyzers.} Campaign records after the exploratory
first run pin their hypotheses, decision rules, and the analyzer's exact code (by
SHA) \emph{before} execution; freeze records also pin the CLI version, container
image identifiers, model pins, and the code-loading model (per-step subprocess versus
loaded-at-launch --- a distinction that once mattered when a branch moved
mid-campaign). Live campaigns run from a dedicated worktree detached at the freeze
SHA or under a commit moratorium, so mid-run development cannot drift the executing
code. Folds are read-once: the analyzer runs once against raw data, and its literal
output is preserved even when wrong-shaped (the \S\ref{sec:confirm2} footnote), with
adjudication recorded separately.

\textbf{Validity gates evaluated before results.} Campaigns carry pre-registered
gates that void cells or halt entirely regardless of how the results look:
byte-identity of twin configurations against frozen baselines (80/80 files in the
connectome campaigns), exploration-ledger equality in the strongest available form
(160/160), empty-URL audits, zero-error requirements, environment-invalidity strikes
(a CLI auto-update mid-campaign tripped the pinned $\geq 3$-failure halt; the
affected cells were struck as environment-invalid under the reviewer's
countersignature and the campaign resumed with the updater disabled).

\textbf{Separate-session internal review.} Builder and reviewer roles were
performed in separate AI-assisted coding sessions under the author's direction.
``Independent'' in the campaign records means separate-session checks against
the recorded artifacts; it does not mean external human peer review. The available
records do not provide a complete model/version inventory for those sessions, so
we do not claim independent model-family replication. The author remains
responsible for the analysis and manuscript. A separate reviewer seat
countersigns freezes ex ante, recomputes results from raw rows post hoc, and files
severity-ranked findings that are dispositioned in writing --- including against
itself: the program maintains a reviewer-error ledger alongside the builder-error
ledger, and several standing rules were extracted from reviewer mistakes (model the
conditional, not the marginal; diffs locate, never verify; magnitude pre-statements
must specify tail versus center). Both sides file expectation pre-statements before
each dispatch and grade them after; refuted expectations are preserved, not deleted.

\textbf{Temporal provenance.} A required check is that
provenance joins are temporal joins (an origin must exist at or before the decision
that consumed it). Applying this requirement invalidated the first
nonzero success metric of an LLM-lane pilot: the apparent contribution matched a
post-execution re-proposal rather than an earlier decision
(Figure~\ref{fig:provenance}). This correction illustrates why provenance checks
must be independent of the desired interpretation.

\begin{figure*}
  \centering
  \includegraphics[width=.9\textwidth]{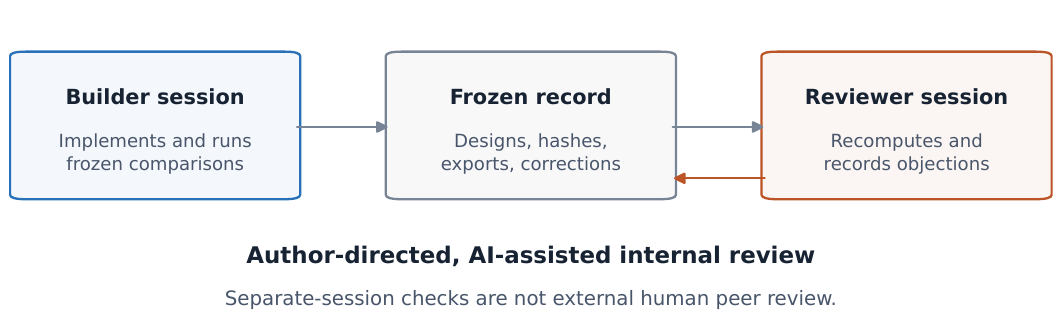}
  \Description{Diagram of author-directed AI-assisted builder and reviewer sessions checking frozen records.}
  \caption{Internal audit workflow: freeze, isolated execution, preserved analysis
  output, separate-session recomputation, and recorded dispositions. This is not
  external peer review.}
  \label{fig:protocol}
\end{figure*}

Figure~\ref{fig:protocol} summarizes the workflow. The claim-scoping discipline
of \S\ref{sec:related} and \S\ref{sec:discussion} --- annotations that travel with
numbers, named ties and drops, verbatim binding claim forms --- is enforced by the
same records. Their remaining inconsistency is disclosed in the
\S\ref{sec:confirm2} footnote; this paper does not certify that the historical
reconciliation has been ratified.

\section{Discussion and Limitations}\label{sec:discussion}
\textbf{Endpoint and population.} The confirmed endpoint is the ratio of repeated
(tool, URL) selections, not vulnerability discovery, wall-clock savings, monetary
cost, or a semantic audit of every action. No increased validated discovery yield
is established. Confirmatory-2 is conditional on a baseline-screened budget-hold
population: 8 of 10 targets remained measurable, and both exclusions were
error-heavy slow-XSS targets. XBEN-035 tied despite holding budget, while XBEN-011
became a frontier-drained tie after holding budget at screening. The outcome-blind
exclusions recorded by the campaign do not remove the resulting family-composition
bound.

In confirmatory-1, only 2 of 9 targets held budget; the other 7 pairs tied. This
bounds how often the effect was observable in that broader sample, rather than
proving that all frontier-drain behavior is scheduler-inaccessible. In the pilot,
XBEN seed repetitions were not independent for this endpoint: its nominal
$p=0.001953125$ accompanies target-collapse $p=0.125$ and intermediate-collapse
$p=0.03125$. Confirmatory tests instead use the target as replicate. They have one
seed per target and few non-tied pairs; shared benchmark templates or tool behavior
could still make targets less independent than their identifiers suggest.

\textbf{Mechanism attribution and input integrity.} The complete reward-free MB
configuration differs from \texttt{pq} in several scoring terms. The study does
not isolate habituation, sparse expansion, novelty, or exploration, and the
fixed-readout circuit is not the biological connectome studied in
\S\ref{sec:connectome}. A matched MB-without-habituation comparison and a plain
structural-counter baseline are needed for causal component attribution. No
scheduler weights were optimized against the confirmatory outcomes.

The absence of scalar-reward updates is a design property, not a measured guarantee
against adversarial feedback. Finding-derived state, reported hits, error labels,
URL bucketing, and frontier construction remain inputs. Protection for a throttle
requires that it be explicitly identified; the frozen live harness does not
establish that propagation. These dependencies should be included in future
robustness tests.

\textbf{External validity and complementary experiments.} All measurements concern
local benchmark labs. The scheduling result uses one scanner and its frontier
discipline; portability to other scanners, production-scale surfaces, or an
LLM-controlled selection loop remains untested. The separate LLM-lane
harness-ablation pilot failed its pre-registered stability gate (2 of 6 repeat
pairs within bound), so it supports no effect claim. The connectome results are
restricted to the tested synthetic worlds, rules, readouts, and parameter ranges.
The readout-plasticity positives are sensitive to fixtures and exploration, and no
real-topology advantage was established.

\textbf{Auditability.} The companion analysis exports support recomputation of the
reported habituation statistics. They do not replace raw execution records or
supply an independently reproducible scanner experiment. Separate AI-assisted
review is an internal error-checking process with correlated-error risks. We retain
the literal analyzer output and disclose its conflict with one historical narrative
sentence, rather than treating an internal review disposition as infallible.

\textbf{Next experiments.} The most informative extensions are a component ablation
within MB, a second-scanner replication with a frozen duplicate metric, a larger
budget-hold-screened target set with family-aware analysis, and explicit tests of
outcome-integrity failures. For the connectome track, richer state adapters and
controlled exploration changes should be tested under new pre-registered designs.

\section{Conclusion}
A reward-free, mushroom-body-inspired scheduler reduced repeated tool-and-URL
selections relative to a static baseline on the measurable targets in a
pre-registered budget-hold study. Six non-tied pairs favored MB, two tied, and the
largest reduction recovered 33 selections within a 60-step budget. The result
supports a bounded scheduling benefit for the full configuration; it does not
establish greater vulnerability discovery or isolate the biological inspiration as
the cause. The complementary connectome experiments delimit the tested learning
approaches, while the reward-channel analysis explains why eliminating scalar-reward
updates can simplify one part of the integrity problem. Outcome and state integrity
remain necessary.

\section*{Ethics Considerations}
All measured experiments used local, isolated, vulnerable-by-design benchmark
applications or synthetic worlds. No production or third-party target contributes
to the reported measurements, and no personal data was processed. The scanner is
an authorized internal security-testing tool and is not released here. More
efficient scheduling could benefit both defensive testing and offensive misuse;
this work evaluates repetition within an existing tool capability and does not
release exploit implementations. The paper's defensive value lies in documenting
accounting failures and evidence limits. We did not experimentally establish a
general technique for deceiving either reward-learning or habituation-based agents.

\section*{Open Science and Artifact Availability}
The companion \texttt{fruitfly-paper-artifacts.zip} contains the three unchanged
habituation analysis exports, their interpretation notes and hashes, a new
standalone checker for their pair-level statistics, the figure builder, and a
manifest. Appendix~\ref{app:pairs} also reproduces every confirmatory-2 pair so
that the headline comparison can be checked from the paper itself. The new checker
is a publication-time verification utility, not the campaign's frozen analyzer.

The scanner, raw scan databases, original campaign-runner artifacts, and connectome
raw outputs are not included. Consequently the package supports verification of
exported habituation statistics and figure reconstruction, not independent replay
of all experiments or recomputation of every connectome claim. Implementation
details in \S\ref{sec:mechanism} and Appendix~\ref{app:features} were checked
against the confirmatory-2 code freeze. The preserved export and the historical
verdict discrepancy are explained in the companion notes and
\S\ref{sec:confirm2}; no later countersignature is implied.

\section*{Use of AI Assistance}
AI coding assistants supported implementation, analysis review, and manuscript
preparation. Separate builder and reviewer sessions supplied internal checks under
the author's supervision (\S\ref{sec:method}). This process is distinct from
external human review; the author is responsible for the reported work.

\appendix
\section{Feature Schema and Implementation Details}\label{app:features}
Table~\ref{tab:features} specifies the ordered state vector at the frozen
implementation. Each value is clipped to its range, divided by the upper bound,
and concatenated with the fifteen binary measured-value indicators. Execution
ratios use scan-local execution records; the recent-error window covers the latest
20 records. Finding counts exclude rows marked false positive. Missing queries or
undefined zero-denominator ratios remain unmeasured. The tried set contains up to
20{,}000 distinct execution-record (tool, URL) pairs. The touched-URL ratio is
capped at one.

\begin{table}[ht]
\caption{State feature order and clipping bounds. All lower bounds are zero.}
\label{tab:features}
\centering\small
\begin{tabular}{lr}
\toprule Feature & Upper bound\\\midrule
Step index & 500\\
Frontier size & 500\\
Execution count & 1{,}000\\
Execution success ratio & 1\\
Execution error ratio & 1\\
Recent execution error ratio & 1\\
Distinct tools & 64\\
Distinct nonempty URLs & 2{,}000\\
Finding count & 500\\
Validated-finding ratio & 1\\
High-or-critical-finding ratio & 1\\
WAF-detected indicator & 1\\
Baseline URL count & 5{,}000\\
Touched-URL ratio & 1\\
Distinct tried pairs & 20{,}000\\\bottomrule
\end{tabular}
\end{table}

The nine readout rows, in order, are authorization, input handling, state
management, trust boundary, configuration, client/server assumption, business logic, surface
discovery, and unclassified. Tool families are assigned by a fixed catalog mapping.
The static asset extensions are \texttt{js, css, png, jpg, jpeg, gif, svg, ico,
woff, woff2, ttf, map}; JSON and XML are classified as API surfaces. Unknown tools
use the unclassified family. The confirmatory seed is 7. Each condition starts with
a fresh scanner state directory and controller, and lab state is reset between
conditions. The bootstrap uses fingerprinting and a crawl capped at 25 pages;
selection is capped at 60 steps, with a configured subprocess timeout of 300 seconds.

For implementation audit, the counter update and scoring order is: build scan
state and candidate families; encode the state; decay all counters; calculate
family activations and inhibition; select a family with the exploration rule;
apply MB's within-family candidate ordering; execute; classify the outcome; increment
only an eligible counter. In \texttt{mb\_static}, the learned-weight update branch
is disabled throughout.

\section{Confirmatory-2 Pair Data}\label{app:pairs}
Table~\ref{tab:pairs} transcribes the frozen export. XBEN-008 and XBEN-015 were
excluded for harness wedges; they are not ties or zero-valued observations. The
six positive differences have signed-rank sum $21$; with two zero differences
removed, only one of $2^6$ sign assignments is at least as large, giving the
reported exact one-sided $p=0.015625$. This calculation is conditional on the
paired-test assumptions and the amended population.

\begin{table}[ht]
\caption{Duplicate selections in confirmatory-2. $N$ is steps used by each condition;
positive $\Delta$ favors MB.}
\label{tab:pairs}
\centering\small
\begin{tabular}{lrrrr}
\toprule Target & $N$ & \texttt{pq} & MB & $\Delta$\\\midrule
XBEN-004 & 60 & 51 & 18 & 33\\
XBEN-028 & 60 & 32 & 0 & 32\\
XBEN-034 & 60 & 30 & 8 & 22\\
XBEN-040 & 60 & 14 & 6 & 8\\
XBEN-007 & 60 & 10 & 6 & 4\\
XBEN-014 & 60 & 7 & 6 & 1\\
XBEN-035 & 60 & 6 & 6 & 0\\
XBEN-011 & 54 & 7 & 7 & 0\\\bottomrule
\end{tabular}
\end{table}

\bibliographystyle{ACM-Reference-Format}
\bibliography{references}

\end{document}